\documentclass[11pt,letterpaper]{article}

\usepackage[margin=1.1in]{geometry}

\usepackage{amsmath,amssymb,amsfonts,setspace,url,enumitem}

\usepackage{amsthm}

\usepackage{tikz}
\usepackage{color}
\usepackage{bigstrut}
\usepackage{multirow}
\usepackage{ascmac}
\usepackage{mathpazo}
\usepackage{authblk}

\makeatletter
\newcommand{\newreptheorem}[2]{\newtheorem*{rep@#1}{\rep@title}\newenvironment{rep#1}[1]{\def\rep@title{#2 \ref*{##1}}\begin{rep@#1}}{\end{rep@#1}}}
\makeatother

\makeatletter
\newcommand{\newconjecture}[2]{\newconjecture*{rep@#1}{\rep@title}\newenvironment{rep#1}[1]{\def\rep@title{#2 \ref*{##1}}\begin{rep@#1}}{\end{rep@#1}}}
\makeatother
\usepackage{comment}
\usepackage{amsthm}
\usepackage{quantikz}
\usepackage{tikz}
\usepackage{mathabx}

\newcommand{\propref}[1]{\hyperref[#1]{Proposition~\ref{#1}}}
\newcommand{\corref}[1]{\hyperref[#1]{Corollary~\ref{#1}}}
\newcommand{\defref}[1]{\hyperref[#1]{Definition~\ref{#1}}}
\newcommand{\secref}[1]{\hyperref[#1]{Section~\ref{#1}}}
\newcommand{\appref}[1]{\hyperref[#1]{Appendix~\ref{#1}}}
\newcommand{\chapref}[1]{\hyperref[#1]{Chapter~\ref{#1}}}
\newcommand{\thmref}[1]{\hyperref[#1]{Theorem~\ref{#1}}}
\newcommand{\lemref}[1]{\hyperref[#1]{Lemma~\ref{#1}}}
\newcommand{\figref}[1]{\hyperref[#1]{Fig.~\ref{#1}}}
\renewcommand{\eqref}[1]{\hyperref[#1]{Eq. (\ref{#1})}}
\newcommand{\ineqref}[1]{\hyperref[#1]{Inequality (\ref{#1})}}
\newcommand{\tableref}[1]{\hyperref[#1]{Table (\ref{#1})}}
\newcommand{\algref}[1]{\hyperref[#1]{Algorithm (\ref{#1})}}
\renewcommand{\epsilon}{\varepsilon}

\newtheorem{innercustomgeneric}{\customgenericname}
\providecommand{\customgenericname}{}
\newcommand{\newcustomtheorem}[2]{%
  \newenvironment{#1}[1]
  {%
   \renewcommand\customgenericname{#2}%
   \renewcommand\theinnercustomgeneric{##1}%
   \innercustomgeneric
  }
  {\endinnercustomgeneric}
}

\newcustomtheorem{customthm}{Theorem}
\newcustomtheorem{customlemma}{Lemma}

\usepackage{booktabs}
\usepackage{algorithm}
\usepackage{amssymb}
\usepackage{amsmath}
\usepackage[colorlinks=true, urlcolor=blue, citecolor=blue]{hyperref}

\usepackage{xcolor}
\definecolor{mina}{rgb}{.2,.5,.1}

\definecolor{elham}{rgb}{.5,.1,.5}

\definecolor{armando}{rgb}{.2,.1,.5}

\usepackage[numbers]{natbib}

\theoremstyle{plain}
\newtheorem{theorem}{Theorem}[section]
\newreptheorem{theorem}{Theorem}
\newreptheorem{conjecture}{Conjecture}
\newtheorem{lemma}{Lemma}[section]

\newtheorem{cor}{Corollary}[section]

\newtheorem{definition}{Definition}[section]

\newtheorem{prop}{Proposition}[section]

\theoremstyle{definition}

\newcommand{\stat}[1]{\vert#1\vert_{\mathrm{tv}}} 

\newcommand{\Tr}{\mathrm{Tr}} 

\newcommand{\tr}{\mathrm{tr}}

\newcommand{\poly}{\mathrm{poly}}

\begin{document}

\title{Quantum Local Differential Privacy and Quantum Statistical Query Model}
\author[,1]{Armando Angrisani \thanks{corresponding author: armando.angrisani@lip6.fr}}
\author[1,2]{Elham Kashefi}

\affil[1]{LIP6, CNRS, Sorbonne Université, 75005 Paris, France}
\affil[2]{School of Informatics, University of Edinburgh, EH8 9AB Edinburgh, United Kingdom}

\date{\today}
\maketitle
\thispagestyle{empty}
\setcounter{page}{1}
\begin{abstract}
Quantum statistical queries provide a theoretical framework for investigating the computational power of a learner with limited quantum resources. This model is particularly relevant in the current context, where available quantum devices are subject to severe noise and have limited quantum memory. On the other hand, the framework of quantum differential privacy demonstrates that noise can, in some cases, benefit the computation, enhancing robustness and statistical security.
In this work, we establish an equivalence between quantum statistical queries and quantum differential privacy in the local model, extending a celebrated classical result to the quantum setting. 
Furthermore, we derive strong data processing inequalities for the quantum relative entropy under local differential privacy and apply this result to the task of asymmetric hypothesis testing with restricted measurements. Finally, we consider the task of quantum multi-party computation under local differential privacy. As a proof of principle, we demonstrate that the parity function is efficiently learnable in this model, whereas the corresponding classical task requires exponentially many samples.

\end{abstract}

\section{Introduction}
Quantum technology has the potential to revolutionize various scientific fields by providing new tools for efficient processing of quantum and classical information. Some of the most promising applications include quantum simulation \cite{daley2022practical}, cryptanalysis \cite{shor1997}, combinatorial optimization \cite{farhi2014quantum}, and topological data analysis \cite{gyurik2022towards}. However, the implementation of many quantum algorithms requires a fault-tolerant quantum computer, which is unlikely to be available in the near future. Therefore, it is crucial to develop realistic models of quantum computation that account for the limitations of near-term hardware \cite{preskill2018quantum}. Recent research has revealed that noise can prevent quantum advantage for specific tasks \cite{stilck2021limitations, de2023limitations, wang2021noise}, or even allow classical computers to efficiently sample from the output distribution of a quantum circuit measured in the computational basis \cite{aharonov2022polynomial}. Error mitigation techniques can improve near-term devices in practical settings, although the recent work of \citet{em} shows that mitigating noise is hard even at a depth of $\poly\log\log (n)$ in the worst-case. Additionally, many quantum algorithms require quantum data as input \cite{anshu2023survey, aaronson2018shadow, huang2020predicting, rouze2021learning}, which may also be corrupted by noise, making a thorough understanding of quantum noise fundamental, even in the presence of fault-tolerant quantum computers.
Motivated by the aforementioned limitations, \citet{arunachalam2020quantum, arunachalam2023role} introduced the model of \emph{quantum statistical query} (QSQ) learning, where a learner can estimate the expectation values of $\poly(n)$-many efficiently implementable \emph{noisy} measurements on (fresh) copies of an unknown input state $\rho$. 
Interestingly, the concept classes of parities, juntas, DNF formulae are efficiently learnable in the QSQ model, whereas they require exponentially many samples in the (classical) statistical query model. Besides, an exponential separation between QSQ learning and learning with quantum examples under classification noise was provided in \cite{arunachalam2023role}.
Quantum statistical queries also found applications in classical verification of quantum learning \cite{caro2023classical}, as well as in the analysis of quantum error mitigation models \cite{em,arunachalam2023role} and quantum neural networks \cite{Du_2021}. Alternative notions of quantum statistical queries were also adopted in \cite{hinsche2022single, gollakota2022hardness, nietner2023average}.

Shifting the focus, a vast body of literature suggests that noise can offer notable benefits for specific computational tasks. Particularly,  noise holds the potential to ensure diverse notions of statistical security, thereby enhancing adversarial robustness and generalization in various settings \cite{chaudhuri2011differentially, dwork2018privacy, cohen2019certified, lecuyer2019certified}. 
In this context, the comprehensive framework of differential privacy emerges as a unifying approach for understanding the role of noise in machine learning and statistics  \cite{dwork1, dwork2, cummings2023challenges}.
Differential privacy comes in several flavors. Particularly, a distinction arises between standard differential privacy and local differential privacy. Notably, the latter model offers a more robust notion of security, as it treats even the curator (i.e., the analyst who accesses the raw input data) as untrusted \cite{equiv,minmax,asoodeh2021local}.
Interestingly, statistical queries and differential privacy are intimately connected: \citet{equiv} showed that a concept class is learnable under (classical) local differential privacy if and only if it is learnable with (classical) statistical queries. 
Recently, several works explored various notions of quantum differential privacy \cite{aaronson2019gentle, quantumZhou, hirche2023quantum, angrisani2023unifying}, finding applications in shadow tomography \cite{aaronson2019gentle}, adversarial classification \cite{Du_2021} and quantum online learning \cite{arunachalam2021private}. Quantum mechanisms for ensuring local differential privacy of classical data were investigated in \cite{yoshida2020classical, yoshida2021mathematical}. 
Locally differentially private (LDP) measurements were introduced in \cite{aaronson2019gentle} and referred as \emph{nearly trivial} measurements. 
Informally, the output of a LDP measurement weakly depends on the input state, and this is often ensured by the injection of noise. This comes with desirable privacy guarantees, along with an increased sample complexity for many computational tasks. Throughout this paper we will argue that certain computational tasks are unfeasible under this strict notion of privacy, while others can be efficiently performed. Crucially, we will demonstrate that local differential privacy is compatible with exponential quantum speed-up for specific tasks.

\paragraph{Our contribution.}
Our first set of contributions consists in several entropic inequalities for locally differentially private channels (\secref{sec:entropic}). In particular, we provide a strong data processing inequality for the quantum relative entropy under locally differentially private measurements. 
In \secref{sec:equiv}, we provide quantum generalization of the equivalence between learning under local differential privacy and statistical query learning, answering an open question posed by \citet{arunachalam2021private}.
As a corollary, we also obtain an exponential separation between learning under quantum local differential privacy and learning with separable measurements, resolving an open question posed by \citet{aaronson2019gentle}.
Furthermore, in \secref{sec:hp}, we provide an application of the aforementioned entropic inequalities to the task of asymmetric hypothesis testing with restricted measurements. Our result is a quantum analog of the private Stein's lemma (\cite{asoodeh2021local}, Corollary 4).
Finally, in \secref{sec:parity} we investigate the problem of learning from quantum data in a distributed setting under local differential privacy. We demonstrate that parity functions are efficiently learnable in this model, whereas the corresponding classical task requires exponentially many samples \cite{equiv}.

\paragraph{Related work.}
The task of quantum hypothesis testing under local differential privacy has also been recently explored in \cite{hirche2023divergence}. We emphasize that \thmref{thm:stein} provides a quadratic improvement over (\cite{hirche2023divergence}, Corollary 5.14) for small values of the privacy level $\epsilon$. It is also worth noting that the results in \cite{hirche2023divergence} extend beyond measurements to encompass private quantum channels. 

\section{Preliminaries}
We start by introducing some preliminary notions of quantum information theory, as well as the basic notions of quantum local differential privacy, quantum statistical queries and Fourier analysis of Boolean functions.
Throughout the paper, we let $[n] = \{1,\dots,n\}$ and for $s \in \{0,1\}^n$, we define $\mathrm{supp}(s) = \{i \in [n] : s_i = 1\}$. We denote by $\mathsf{conv}(X)$ the convex hull of the set $X$, i.e. the set of all convex combinations of the elements in $X$.

\subsection{Quantum information theory}
 We briefly review the basic concepts in quantum information theory. We define $\ket{0}:= (1\;\;0)^\intercal$ and $\ket{1}:= (0\;\;1)^\intercal$ 
as the canonical basis for $\mathbb{C}^2$. An $n$-qubit pure quantum state $\ket{\psi}$ is a unit vector in $\mathbb{C}^{2^n}$ and can be expressed as $\ket{\psi}=\sum_{x \in \{0,1\}^n} \alpha_x \ket{x}$ where $\alpha_x \in \mathbb{C}$ and $\sum_x
|\alpha_x|^2 = 1$. We denote by $\bra{\psi}$ as the conjugate transpose of the quantum state $\ket{\psi}$. 
In general, we may also have classical probability distributions over pure states. This scenario is captured by \emph{mixed} states, the most general kind of states in quantum mechanics. Mixed states are described by density matrices. Formally, a $d$-dimensional mixed state $\rho$ is a $d \times d$ positive semidefinite matrix that satisfies $\text{Tr}(\rho)=1$. 
A quantum channel $\mathcal{C}$ is a completely positive
and trace-preserving linear map, which maps a mixed state $\rho$ to the mixed state $\mathcal{C}(\rho) = \sum_{i=1}^k B_i \rho B_i^\dag$,
where $B_1,\ldots,B_k$ can be any matrices satisfying $\sum_{i=1}^k B_i^\dag B_i = \mathbb{1}$.
The most general class of measurements that we can perform on mixed states are the POVM (Positive Operator Valued Measure) measurements. Although they can be represented as channels, it's convenient to define them separately. In the POVM formalism, a measurement $\mathcal{M}$ is given by a list of $d\times d$ positive semidefinite matrices $(\mathcal{M}_1,\ldots, \mathcal{M}_{k})$, which
satisfy $\sum_{i=1}^k \mathcal{M}_i = \mathbb{1}$. Each $\mathcal{M}_i$ is called POVM element. The measurement rule is:
\[
\Pr[\mathcal{M}\text{ returns outcome }i\text{ on input }\rho]=\Tr(\mathcal{M}_i\rho).
\]
We'll denote as $\mathcal{M}(\rho)$ the distribution over $[k]$ induced by performing $\mathcal{M}$ on the state $\rho$. Thus we have $\mathbb{E}[\mathcal{M}(\rho)]=\sum_{i=1}^k i \cdot \Tr(\mathcal{M}_i\rho)$.
Given a Hermitian matrix $A$ with eigenvalues $\lambda_1,\dots,\lambda_d$, its trace norm is defined as
$\|A\|_\tr:= \Tr|A| = \frac{1}{2}\sum_{i=1}^d \lambda_i$.
Moreover, the trace distance between two mixed states $\rho$ and $\sigma$ is defined as $\|\rho - \sigma\|_\tr$.
It's convenient to write the following spectral decomposition 
\[
\rho-\sigma = \sum_{i} \lambda_i \ket{i}\bra{i} = X^+ - X^-, 
\]
where $X^+$ and $X^-$ denote respectively the positive part and the negative part of $\rho-\sigma$, i.e.
\[
X^+ := \sum_{\lambda_i>0} \lambda_i \ket{i}\bra{i}, \;\;\; X^- := \sum_{\lambda_i<0} \lambda_i \ket{i}\bra{i}.
\]
In particular, the following identities can be easily verified:
\begin{equation}
    \|\rho-\sigma\|_\tr = \frac{1}{2}\Tr(X^+) = \frac{1}{2}\Tr(X^-). 
\end{equation}
We also introduce several quantum divergences. For two states $\rho, \sigma$
such that the support of $\rho$ is included in the support of $\sigma$, the quantum relative entropy is defined as
\[D(\rho\|\sigma) = \Tr[\rho (\log \rho -\log \sigma)].\]
The quantum relative entropy can be lower bounded by the \emph{measured} relative entropy, defined as follows:
\[
D_M(\rho\|\sigma) = \sup_{\mathcal{M}} D(\mathcal{M}(\rho)\|\mathcal{M}(\sigma)),
\]
where the supremum is taken over all POVM measurements.
Under the same assumption on the supports of $\rho$ and $\sigma$, the quantum max-relative entropy \cite{datta2009min} is defined as
\[
D_{\max}(\rho\|\sigma) = \inf\{\lambda : \rho \leq e^\lambda \sigma\}.
\]
The quantum smooth max-relative entropy \cite{hirche2023quantum} is a relaxation of the quantum max-relative entropy defined as follows:
\[
D_{\max}^\delta (\rho\|\sigma)= \inf_{\overline{\rho}\in B_\delta(\rho)} D_{\max}(\overline{\rho}\|\sigma),
\]
where $B^\delta (\rho) = \{\overline{\rho} : \overline{\rho}^\dag = \overline{\rho}\geq 0 \wedge \|\rho-\overline{\rho}\|_1<2\delta\}$.
We will also need the quantum hockey stick divergence of order $\gamma\geq 1$ \cite{sharma2012strong}, which is defined as 
\[
E_\gamma(\rho\|\sigma)= \Tr(\rho-\gamma \sigma)^+.
\]
We remark that $E_1(\rho\|\sigma):= \|\rho-\sigma\|_\tr$.

\subsection{Quantum local differential privacy}
\label{sec:qdp}
A quantum channel $\mathcal{N}$ is $(\epsilon,\delta)$-locally differentially private (LDP) if for every POVM measurement $\mathcal{M}=\{\mathcal{M}_x\}_{x\in\mathcal{X}}$ and for all pairs of states $\rho,\sigma$,
\begin{equation}
\label{eq:ldp}
   \forall x\in \mathcal{X} : \Tr[\mathcal{M}_x \mathcal{N}(\rho) ]\leq e^\epsilon \Tr[\mathcal{M}_x \mathcal{N}(\sigma)] +\delta. 
\end{equation}

As shown in \cite{hirche2023quantum}, quantum differential privacy can be equivalently expressed in terms of  the {quantum hockey-stick divergence} and the {quantum smooth max-relative entropy}:
\begin{equation*}
   \mathcal{N} \text{ is $(\epsilon,\delta)$-LDP } \iff \forall \rho,\sigma: D^\delta_{\max}(\mathcal{N}(\rho)\|\mathcal{N}(\sigma))\leq \epsilon \iff \forall \rho,\sigma: E_{e^\epsilon}(\mathcal{N}(\rho)\|\mathcal{N}(\sigma))\leq \delta. 
\end{equation*}

The special case where $\delta = 0$ is usually referred as \emph{pure} local differential privacy, while the most general case is referred as \emph{approximate} local differential privacy. When $\delta = 0$, we will write $\epsilon$-LDP instead of $(\epsilon,0)$-LDP.
For the special case of quantum-to-classical channel, i.e. POVM measurements, local differential privacy is also referred as \emph{near triviality}  \cite{aaronson2019gentle}. In particular, we say that a POVM measurement $\mathcal{M}=\{\mathcal{M}_x\}_{x\in\mathcal{X}}$ is $(\epsilon,\delta)$-LDP -- or $(\epsilon,\delta)$-trivial -- if for all states $\rho,\sigma$ and for all $F\subseteq \mathcal{X}$,
\[
\Pr[\mathcal{M}(\rho)\in F] \leq e^\epsilon \Pr[\mathcal{M}(\sigma)\in F] + \delta.
\]
It's easy to see that if $\mathcal{N}$ is an $(\epsilon,\delta)$-LDP channel, then for all measurement $\mathcal{M}$, the measurement $\mathcal{M(\mathcal{N(\cdot)})}$ is also $(\epsilon,\delta)$-LDP. This can be proven by invoking the monotonicity of the quantum hockey-stick divergence.
As shown in \cite{aaronson2019gentle}, all $\epsilon$-LDP measurements admit a $\epsilon$-gentle implementation, i.e. an implementation that produces a post-measurement state which is $O(\epsilon)$-close in trace distance to the input state. 
We also remark that the notion of local differential privacy can be relaxed to (standard) differential privacy, by requiring the inequalities in \eqref{eq:ldp} only if $\rho$ and $\sigma$ satisfy some suitable ``neighboring relationship''. However, an introduction to differential privacy beyond the local model exceeds the scope of this paper, hence we refer to \cite{quantumZhou, aaronson2019gentle, hirche2023quantum, angrisani2023unifying} for more details.
We will now present two mechanisms mapping any measurement to a suitable LDP measurement. 

\paragraph{Quantum randomized response.}
We introduce a quantum version of the \emph{randomized response} mechanism, which is widely used in (classical) local differential privacy \cite{rr,kairouz2014extremal}.
Given an arbitrary POVM measurement $\mathcal{M} = (\mathcal{M}_1,\mathcal{M}_2,\dots, \mathcal{M}_k)$,  we define $\mathcal{M}^{\mathsf{RR},\epsilon} = (\mathcal{M}^{\mathsf{RR},\epsilon}_1,\mathcal{M}^{\mathsf{RR},\epsilon}_2,\dots, \mathcal{M}^{\mathsf{RR},\epsilon}_k)$ as follows:
\[
\mathcal{M}_i^{\mathsf{RR}, \epsilon} = \frac{e^\epsilon-1}{e^\epsilon - 1 + k} \mathcal{M}_i + \frac{1}{e^\epsilon - 1 + k}\mathbb{1}.
\]
It's easy to verify that $\mathcal{M}^{\mathsf{RR}, \epsilon}$ satisfies $\epsilon$-local differential privacy.
For an arbitrary state $\rho$, we have
\[
\frac{1}{e^\epsilon - 1 + k}\leq \Tr[\mathcal{M}_i^{\mathsf{RR},\epsilon} \rho] \leq \frac{e^\epsilon}{e^\epsilon - 1 + k}
\]
and thus for all states $\rho,\sigma$, 
\[\frac{\Tr[\mathcal{M}_i^{\mathsf{RR},\epsilon} \rho]}{\Tr[\mathcal{M}_i^{\mathsf{RR},\epsilon} \sigma]} \leq e^\epsilon.\]
In expectation, the quantum randomized response yields:
\[
\mathbb{E}\left[\mathcal{M}^{\mathsf{RR},\epsilon}(\rho)\right] = \frac{e^\epsilon-1}{e^\epsilon - 1 + k}\mathbb{E}\left[\mathcal{M}(\rho)\right] + \frac{k}{2(e^\epsilon - 1 + k)}, 
\]
so $\mathcal{M}^{\mathsf{RR},\epsilon}(\rho)$ is a biased estimator for $\mathbb{E}\left[\mathcal{M}(\rho)\right]$.
\paragraph{Quantum Laplace measurement.}
We now recall the definition of the \emph{Laplace measurement}, introduced in \cite{aaronson2019gentle} and inspired by the classical Laplace mechanism \cite{dwork1,dwork2}.
Given an input state $\rho$ and an arbitrary $k$-ary POVM measurement $\mathcal{M}=\{\mathcal{M}_1,\mathcal{M}_2,\dots,\mathcal{M}_k\}$, the Laplace measurement $\mathcal{M}^{\mathsf{Lap},\epsilon}$ associated to $\mathcal{M}$ can be implemented by sampling $y\sim \mathcal{M}(\rho)$  and then releasing $y+\eta$, where $\eta$ is sampled from the Laplace distribution centered in $0$ and with scale parameter $(k-1)/\epsilon$:
\[
\eta \sim \frac{\epsilon}{2(k-1)} \exp\left(-\frac{\epsilon}{k-1}|\eta|\right).
\]
Note that the Laplace measurement has range $\mathbb{R}$, even if the underlying measurement $\mathcal{M}$ has discrete range.
Let $\hat{\mu} = y +\eta$ the output of the Laplace measurement. It's easy to see that $\Pr[\hat{\mu}\text{ on input }\rho]\leq e^\epsilon \Pr[\hat{\mu}\text{ on input }\sigma]$ for all states $\rho$ and $\sigma$, and hence the Laplace measurement satisfies $\epsilon$-local differential privacy. Moreover, $\mathbb{E}\hat{\mu} = \mathbb{E} [\mathcal{M}(\rho)]$, so $\hat{\mu}$ is an unbiased estimator for $\mathbb{E} [\mathcal{M}(\rho)]$.

\subsection{Quantum statistical queries}
Quantum statistical queries (QSQs) \cite{arunachalam2020quantum} are a quantum extension of \emph{classical} statistical queries, introduced in \cite{sq}.
In the QSQ model, a learner -- that is still a classical randomized algorithm -- can query an oracle to obtain statistics about an unknown quantum state. This framework is motivated by the practical constraints encountered on near-term devices, which are severely affected by noise and dispose of limited quantum memory. 
We now give the definition of quantum statistical query provided in \cite{arunachalam2023role}.
\begin{definition}[QSQ oracle]
Let $\rho$ an unknown (mixed) quantum state. A quantum statistical query oracle $\mathsf{QStat}_\rho(\tau,M)$, receives as input an operator $M$, satisfying $\|M\|\leq 1$, and a tolerance parameter $\tau\geq 0$, and outputs a $\tau$-approximation of $\Tr(M\rho)$,
\[
\mathsf{QStat}_\rho :  (M,\tau) \mapsto \alpha \in [\Tr(M\rho)-\tau, \Tr(M\rho) +\tau].
\]
\end{definition}
A case of particular interest is when $\rho$ is a \emph{quantum example} \cite{bshouty}, i.e. $\rho = \ket{\psi_{f,\mathcal{D}}}\bra{\psi_{f,\mathcal{D}}}$ is a quantum encoding of a classical Boolean function $f:\{0,1\}^n\rightarrow \{0,1\}$ with respect to a distribution $\mathcal{D}:\{0,1\}^n\rightarrow [0,1]$,
\[
\ket{\psi_{f,\mathcal{D}}} = \sum_{x \in \{0,1\}^n} \sqrt{D(x)}\ket{x}\ket{f(x)}.
\]

Quantum examples are the building blocks of quantum \emph{probably approximately correct} (PAC) learning , where a learner is given as input $m$ copies of a state $\ket{\psi_{f,\mathcal{D}}}\bra{\psi_{f,\mathcal{D}}}$ and whose goal is to output a function $\widehat{f}$ satisfying
\[
\mathbb{E}_{x\sim \mathcal{D}} |f(x) - \widehat{f}(x)| \leq \epsilon\;\; \text{with probability at least $1-\delta$},
\]
for some parameters $\epsilon,\delta \in [0,1]$.

\subsection{Fourier analysis}
We now introduce some basic notions of Fourier analysis on the Boolean cube. For a comprehensive introduction to the topic, we refer to \cite{o2021analysis}. For $S\in\{0,1\}^n$, we define the \emph{character function} $\chi_S:\{0,1\}^n\mapsto \{-1,1\}^n$ as $\chi_S=(-1)^{S\cdot x}$, where $S\cdot x = \sum_i s_i \cdot x_i \mod 2$.  Notably, Boolean functions can be decomposed in terms of {character functions}, weighted according to their associated \emph{Fourier coefficients}, which are defined as
\[
\widehat{f}(S) = \mathbb{E}_{x\in\{0,1\}^n} [f(x) \cdot \chi_S(x)],
\]
where the expectation is taken over $x$ sampled uniformly at random from $\{0,1\}^n$. Then every function $f:\{0,1\}^n \mapsto \mathbb{R}$ can be uniquely written as $f(x) = \sum_{S\in\{0,1\}^n} \widehat{f}(S)\chi_S(x)$.
For all $i \in [n]$, we define the $i$-th influence of $f$ as
\[
\mathrm{Inf}_i(f) = \sum_{\substack{S\in\{0,1\}^n: \\ S_i=1}} \widehat{f}(S)^2.
\]
\section{Entropic inequalities under local privacy}
\label{sec:entropic}
 A crucial fact in quantum information theory is that many physical quantities are monotone under the application of a quantum channel. For instance, the quantum relative entropy satisfies the following \emph{data-processing inequality} (DPI), for all states $\rho,\sigma$ and for every channel $\mathcal{N}$:
\[
D(\mathcal{N}(\rho)\|\mathcal{N}(\sigma)) \leq D(\rho\|\sigma).
\]
Furthermore, the same property is shared by the hockey-stick divergences, and in particular by the trace distance. When the inequality is strict, we say that a given divergence satisfies a \emph{strong} data-processing inequality (SDPI) with respect to the channel $\mathcal{N}$.
We can also consider the following contraction coefficients, previously considered in \cite{lesniewski1999monotone,hiai2016contraction,hirche2022contraction,hirche2023quantum}.
\[
\eta(\mathcal{N}) := \sup_{\rho,\sigma} \frac{D(\mathcal{N}(\rho)\|\mathcal{N}(\sigma))}{D(\rho\|\sigma)} \;\;\text{ and }\;\; \eta_{\gamma}(\mathcal{N}) := \sup_{\rho,\sigma} \frac{E_\gamma(\mathcal{N}(\rho)\|\mathcal{N}(\sigma))}{E_\gamma(\rho\|\sigma)}. 
\]
where $\gamma \geq 1$. Recall that $E_1(\rho\|\sigma)= \|\rho-\sigma\|_\tr$ and hence $\eta_1(\mathcal{N}) $ is the contraction coefficient for the trace distance.
If $\mathcal{N}$ satisfies $(\epsilon,\delta)$-LDP, then its contraction coefficient $\eta_\gamma$ can be upper bounded as follows (\cite{hirche2023quantum}, Theorem II.2 and Corollary V.1):
\begin{equation}
\label{eq:phi}
    \eta_{e^\epsilon}(\mathcal{N}) \leq \delta\;\;\text{and}\;\; \eta_\gamma(\mathcal{N}) \leq \varphi(\epsilon,\delta),
\end{equation}
where $\varphi (\epsilon,\delta):= 1- e^{-\epsilon}(1-\delta)$.
More broadly, we can also consider inequalities involving two distinct divergences. For instance, every $\epsilon$-LDP channel $\mathcal{N}$ satisfies:
\begin{equation}
\label{eq:measured}
        D_M(\mathcal{N}(\rho)\|\mathcal{N}(\sigma)) \leq 2 \epsilon \|\mathcal{N}(\rho) - \mathcal{N}(\sigma)\|_\tr
    \leq 2\epsilon (1-e^{-\epsilon}) \|\rho-\sigma\|_\tr,
\end{equation}
where the first inequality is due to (\cite{hirche2023quantum}, Lemma III.6) and the second inequality follows from \eqref{eq:phi}. We will now prove an analogous result, where the measured relative entropy is replaced by the quantum relative entropy.

\begin{prop}
\label{prop:quantum-rel}
For all states $\rho,\sigma$ we have
\[
 D\left(\rho\|\sigma\right) +  D\left(\sigma\|\rho\right) \leq [D_{\max}  (\rho\|\sigma)+D_{\max}(\sigma\|\rho)]\|\rho - \sigma\|_\tr
\]
\end{prop}
\begin{proof}
Recall that we can write the decomposition $\rho-\sigma = X^+ - X^-$, where $X^+$ and $X^-$ denote respectively the positive part and the negative part of $\rho-\sigma$.
We start by rearranging the expression of the quantum relative entropy as follows
\begin{align*}
    D\left(\rho\|\sigma\right) +  D\left(\sigma\|\rho\right)= \Tr\left[\rho\left(\log \rho-\log\sigma\right)\right] + \Tr\left[\sigma\left(\log \sigma-\log\rho\right)\right] 
    \\= \Tr\left[\left(\rho-\sigma\right)\left(\log \rho-\log\sigma\right)\right] = \Tr\left[\left(X^+ - X^-\right)\left(\log \rho-\log\sigma\right)\right] \\= \Tr\left[X^+ \left(\log \rho-\log\sigma\right)\right] + \Tr\left[X^- \left(\log \sigma-\log\rho\right)\right] , 
\end{align*}
By definition of max-relative entropy, $\rho \leq e^{D_{\max} (\rho\|\sigma)} \sigma$. Since the logarithm is an operator monotone function, we have that $\log\rho \leq  \log\left(e^{D_{\max} (\rho\|\sigma)} \sigma\right) = {D_{\max} (\rho\|\sigma)} \mathbb{1} + \log\sigma$. Similarly, we also have $\log\sigma \leq   {D_{\max} (\sigma\|\rho)}\mathbb{1} + \log\rho$.  Putting all together, we obtain
\begin{align*}
    D\left(\rho\|\sigma\right) +  D\left(\sigma\|\rho\right) \leq {D_{\max}  (\rho\|\sigma)} \Tr\left[X^+\right] + {D_{\max}  (\sigma\|\rho)} \Tr\left[X^-\right] 
    \\= [D_{\max}  (\rho\|\sigma)+D_{\max}(\sigma\|\rho)]\|\rho - \sigma\|_\tr,
\end{align*}
where the equality follows from $\Tr\left[X^+\right] = \Tr\left[X^-\right]= \|\rho-\sigma\|_\tr$.
\end{proof}
We remark that an analogous result has also been recently presented in (\cite{hirche2023divergence}, Eqs. 5.25-27). However, in \cite{hirche2023divergence} the sum $D\left(\rho\|\sigma\right) +  D\left(\sigma\|\rho\right)$ is replaced by $D\left(\rho\|\sigma\right)$. Thus, our result is tighter of a factor 2 when the goal is to upper bound the sum $D\left(\rho\|\sigma\right) +  D\left(\sigma\|\rho\right)$.
A simple application of \eqref{eq:phi} to \propref{prop:quantum-rel} yields the following corollary, which generalizes \eqref{eq:measured}.
\begin{cor}
\label{cor:quantum-rel}
Let $\mathcal{N}$ an $\epsilon$-LDP channel. Then for all states $\rho,\sigma$ we have
\[
D\left(\mathcal{N}(\rho)\|\mathcal{N}(\sigma)\right) +  D\left(\mathcal{N}(\sigma)\|\mathcal{N}(\rho)\right) \leq 2 \epsilon (1-e^{-\epsilon})\|\rho - \sigma\|_\tr.
\]
\end{cor}

We now derive yet another improved version of \eqref{eq:measured}, by generalizing Lemma 1 in \cite{minmax} to the quantum setting.  
\begin{lemma}
\label{lem:duchi}
Let $\mathcal{M} = \{\mathcal{M}_x\}_{x\in X}$ be an $\epsilon$-LDP POVM measurement. Then for all states $\rho,\sigma$ we have
\[
D\left(\mathcal{M}(\rho)\|\mathcal{M}(\sigma)\right) +  D\left(\mathcal{M}(\sigma)\|\mathcal{M}(\rho)\right) \leq e^\epsilon (1-e^{-\epsilon}  )^2 \|\rho-\sigma\|_\tr^2.
\]
Moreover, for every $\epsilon$-LDP channel $\mathcal{N}$ and for all states $\rho,\sigma$,
\[
D_M\left(\mathcal{N}(\rho)\|\mathcal{N}(\sigma)\right) \leq e^\epsilon (1-e^{-\epsilon} )^2 \|\rho-\sigma\|_\tr^2,
\]
where $D_M(\cdot\|\cdot)$ is the measured relative entropy.
\end{lemma}
\begin{proof}
Let $p_x = \Tr (\mathcal{M}_x\rho)$ and $q_x = \Tr (\mathcal{M}_x\sigma)$.
\begin{align*}
     D(\mathcal{M}(\rho)\|\mathcal{M}(\sigma))+ D(\mathcal{M}(\sigma)\|\mathcal{M}(\rho))\\= \sum_x p_x \log\frac{p_x}{q_x} + \sum_x q_x \log\frac{q_x}{p_x} = 
    \sum_x (p_x-q_x) \log\frac{p_x}{q_x}.
\end{align*}
We want to upper bound $|p_x-q_x| = |\Tr (\mathcal{M}_x(\rho-\sigma))|$. Let $\rho-\sigma = X^+ - X^-$, where $X^+$ and $X^-$ denote respectively the positive part and the negative part of $\rho-\sigma$. We can also write the spectral decompositions $X^+ = \sum_{y \in \mathcal{Y}} \lambda_y \ket{y}\bra{y}$ and $X^- = \sum_{z \in \mathcal{Z}} \tau_z \ket{z}\bra{z}$.
First, we upper bound $p_x-q_x = \Tr (\mathcal{M}_x(\rho-\sigma))$
\begin{align*}
    \Tr (\mathcal{M}_x(X^+ - X^-)) = \Tr\left(\mathcal{M}_x \left( \sum_{y \in \mathcal{Y}} \lambda_y \ket{y}\bra{y}\right)\right) - \Tr\left(\mathcal{M}_x\left(\sum_{z \in \mathcal{Z}}\tau_z \ket{z}\bra{z}\right)\right) \\
    \leq \max_{y \in \mathcal{Y}} \Tr(\mathcal{M}_x \ket{y}\bra{y}) \left(\sum_{y \in \mathcal{Y}} \lambda_y\right) - \min_{z \in \mathcal{Z}} \Tr(M_x \ket{z}\bra{z}) \left(\sum_{z \in \mathcal{Z}} \tau_z\right) \\
    = \|\rho-\sigma\|_\tr \left(\max_{y \in \mathcal{Y}} \Tr(\mathcal{M}_x \ket{y}\bra{y}) -\min_{z \in \mathcal{Z}} \Tr(\mathcal{M}_x \ket{z}\bra{z})\right)\\
    \leq \|\rho-\sigma\|_\tr \max_{y \in \mathcal{Y}} \Tr(\mathcal{M}_x \ket{y}\bra{y})(1-e^{-\epsilon}),
\end{align*}
where the second equality follows from the identities $\Tr[X^+] = \sum_{y \in \mathcal{Y}} \lambda_y = \|\rho-\sigma\|_\tr$ and $\Tr[X^-] = \sum_{z \in \mathcal{Z}} \tau_z = \|\rho-\sigma\|_\tr$, and the last inequality follows from $\epsilon$-LDP.
Proceeding in an analogous way, we derive the following lower bound.
\begin{align*}
    \Tr (\mathcal{M}_x(X^+ - X^-)) \geq 
    \min_{y \in \mathcal{Y}} \Tr(\mathcal{M}_x \ket{y}\bra{y}) \left(\sum_{y \in \mathcal{Y}} \lambda_y\right) - \max_{z \in \mathcal{Z}} \Tr(\mathcal{M}_x \ket{z}\bra{z}) \left(\sum_{z \in \mathcal{Z}} \tau_z\right) \\
    = \|\rho-\sigma\|_\tr \left(\min_{y \in \mathcal{Y}} \Tr(\mathcal{M}_x \ket{y}\bra{y}) -\max_{z \in \mathcal{Z}} \Tr(\mathcal{M}_x \ket{z}\bra{z})\right)\\
    \geq \|\rho-\sigma\|_\tr \max_{z \in \mathcal{Z}} \Tr(\mathcal{M}_x \ket{z}\bra{z})(e^{-\epsilon} - 1),
\end{align*}
where we applied again the identities $\Tr[X^+] = \Tr[X^-] = \|\rho-\sigma\|_\tr$ and  $\epsilon$-LDP.
We can now provide an upper bound for $|\Tr (\mathcal{M}_x(\rho - \sigma))|$:
\begin{align*}
   |\Tr (\mathcal{M}_x(\rho - \sigma))| = |\Tr(\mathcal{M}_x(X^+-X^-)|
   = \max \{\Tr(\mathcal{M}_x(X^+-X^-), \Tr(\mathcal{M}_x(X^--X^+)\}\\
    \leq  \|\rho-\sigma\|_\tr (1-e^{-\epsilon}) \max_{y\in \mathcal{Y}\cup\mathcal{Z}} \Tr(\mathcal{M}_x \ket{y}\bra{y}),
\end{align*}
Recall that, for $a,b \in \mathbb{R}_+$ (\cite{minmax}, Lemma 4)
\[
\log \frac{a}{b} \leq \frac{|a-b|}{\min\{a,b\}} 
\]
Thus,
\begin{align*}
    \log \frac{p_x}{q_x} \leq \frac{|p_x-q_x|}{\min\{p_x,q_x\}} = \frac{|\Tr (\mathcal{M}_x(\rho - \sigma))|}{\min\{\Tr (\mathcal{M}_x\rho),\Tr (\mathcal{M}_x\sigma)\}}
    \\ \leq \frac{\|\rho-\sigma\|_\tr (1-e^{-\epsilon}) \max_{y\in \mathcal{Y}\cup \mathcal{Z}} \Tr(\mathcal{M}_x \ket{y}\bra{y})}{\min\{\Tr (\mathcal{M}_x\rho),\Tr (\mathcal{M}_x\sigma)\}} \leq  e^\epsilon(1-e^{-\epsilon}) \|\rho-\sigma\|_\tr,
\end{align*}
where we applied $\epsilon$-LDP in the last inequality.
Putting all together,
\begin{align*}
    D(\mathcal{M}(\rho)\|\mathcal{M}(\sigma))+ D(\mathcal{M}(\sigma)\|\mathcal{M}(\rho)) 
    \\\leq e^\epsilon(1-e^{-\epsilon}) \|\rho-\sigma\|_\tr  \left(\|\rho-\sigma\|_\tr (1-e^{-\epsilon}) \max_{y\in \mathcal{Y}\cup \mathcal{Z}} \Tr(\mathcal{M}_x \ket{y}\bra{y})\right) \\
    \leq e^\epsilon (1-e^{-\epsilon})^2 \|\rho-\sigma\|_\tr^2,
\end{align*}
where the last inequality follows from $\max_{y\in \mathcal{Y}\cup \mathcal{Z}} \Tr(\mathcal{M}_x \ket{y}\bra{y}) \leq 1$.
We proved the first part of the lemma. As for the second part, let $\widehat{\mathcal{M}}$ the measurement that maximizes $D(\mathcal{N}(\widehat{\mathcal{M}}(\rho))\|\mathcal{N}(\widehat{\mathcal{M}}(\sigma)))$. Then the desired result follow from the definition of measured relative entropy.
\begin{align*}
    D_M(\mathcal{N}(\rho)\|\mathcal{N}(\sigma)) = D(\mathcal{N}(\widehat{\mathcal{M}}(\rho))\|\mathcal{N}(\widehat{\mathcal{M}}(\sigma))) 
\\ \leq  D(\mathcal{N}(\widehat{\mathcal{M}}(\rho))\|\mathcal{N}(\widehat{\mathcal{M}}(\sigma))) + D(\mathcal{N}(\widehat{\mathcal{M}}(\sigma))\|\mathcal{N}(\widehat{\mathcal{M}}(\rho)))
\\ \leq e^\epsilon (1-e^{-\epsilon} )^2 \|\rho-\sigma\|_\tr^2.
\end{align*}
\end{proof}

We observe that, for small values of $\epsilon$, \lemref{lem:duchi} is quadratically tighter in $\|\rho-\sigma\|_\tr$ with respect to \eqref{eq:measured}. A simple application of the ``measured'' Pinsker's inequality (\lemref{lem:pinsker}) to \lemref{lem:duchi} yields the following corollary.

\begin{cor}
\label{cor:quadratic}
Let $\mathcal{M} = \{\mathcal{M}_x\}_{x\in X}$ be an $\epsilon$-LDP POVM measurement. Then for all states $\rho,\sigma$ we have
\[
D\left(\mathcal{M}(\rho)\|\mathcal{M}(\sigma)\right) +  D\left(\mathcal{M}(\sigma)\|\mathcal{M}(\rho)\right) \leq \frac{e^\epsilon}{2} (1-e^{-\epsilon} )^2 D_M(\rho\|\sigma),
\]
where $D_M(\cdot\|\cdot)$ is the measured relative entropy.
Moreover, for every $\epsilon$-LDP channel $\mathcal{N}$ and for all states $\rho,\sigma$,
\[
D_M\left(\mathcal{N}(\rho)\|\mathcal{N}(\sigma)\right) \leq \frac{e^\epsilon}{2} (1-e^{-\epsilon})^2 D_M(\rho\|\sigma).
\]

\end{cor}
\section{Learning under local privacy is equivalent to QSQ learning}
\label{sec:equiv}
In this section we show an equivalence between locally differentially private measurements and quantum statistical queries, answering an open question posed in (\cite{arunachalam2021private}, Question 7). 
In particular, we will prove that quantum statistical queries can be efficiently simulated by differentially private measurements, and vice versa, differentially private measurements can be efficiently simulated by quantum statistical queries. The latter result is less intuitive and relies on a rejection-sampling argument. The classical analog of the equivalence was proven in the seminal paper of \citet{equiv}.
Interestingly, this result readily implies an exponential separation between learning under quantum local differential privacy and learning with separable measurements, answering an open question posed in (\cite{aaronson2019gentle}, Question 4).

 \subsection{Simulation of QStat queries with locally differentially private measurements}
We first show that quantum statistical queries can be simulated efficiently with LDP measurements. 
The result follows by iterating the Laplace measurement defined in \secref{sec:qdp} and using concentration of measure.
\begin{theorem}
\label{thm:equiv}
If $m\geq c\cdot\frac{\log(1/\beta)k^2}{\epsilon^2 \tau^2}$ for a sufficiently large constant $c$, then $\mathcal{A}_{\mathcal{M},\epsilon}$ (\algref{alg1}) approximates $\mu=\mathbb{E}[\mathcal{M}(\rho)]$ within additive error $\pm \tau$  with probability at least $1-\beta$. 
Moreover, each measurement performed by $\mathcal{A}_{\mathcal{M},\epsilon}$ satisfies $\epsilon$-local differential privacy.
\end{theorem}
\begin{proof}
The proof closely follows the one of Lemma 5.6 in \cite{equiv}.
\algref{alg1} implements the Laplace measurement $\mathcal{M}^{\mathrm{Lap},\epsilon}$ on each copy of $\rho$ and then averages the results.
We first show that $\frac{1}{m}\sum_i y_i$ is concentrated around $\mu:=\mathbb{E}[\mathcal{M}(\rho)]$. By the Chernoff-Hoeffding bound for real-valued variables,
\[
\Pr\left[\left|\frac{1}{m}\sum_{i=1}^m y_i - \mu\right|\geq \frac{\tau}{2}\right] \leq 2 \exp \left(-\frac{\tau^2 m}{2k^2}\right).
\]
The contribution of the Laplace noise can also be bounded via a standard tail inequality. By Lemma A.3 in \cite{equiv},
\[
\Pr\left[\left|\frac{1}{m}\sum_{i=1}^m \eta_i \right|\geq \frac{\tau}{2}\right] \leq \exp\left(- \frac{\tau^2\epsilon^2 m}{4k^2}\right)
\]
And thus by union bound,
\[
\Pr[|\hat{\mu} - \mu|\geq \tau] \leq 2 \exp \left(-\frac{\tau^2 m}{2k^2}\right) + \exp\left(- \frac{\tau^2\epsilon^2 m}{4k^2}\right)\leq 3 \exp \left(-\frac{\tau^2 \epsilon^2 m}{4k^2}\right), 
\]
where $\hat{\mu}:= \frac{1}{m}\sum_{i=1}^m (y_i+\eta_i)$. This implies that $O\left(\frac{\log(1/\beta)k^2}{\epsilon^2 \tau^2}\right)$ samples are sufficient to ensure that $\hat{\mu}$ approximates $\mu$ within additive error $\pm \tau$ with probability at least $1-\beta$.
Moreover, each Laplace measurement $\mathcal{M}^{\mathrm{Lap},\epsilon}$ satisfies $\epsilon$-local differential privacy.
\end{proof}
\thmref{thm:equiv} can be easily extended to the case where an algorithm $\mathcal{B}$ makes $t$ queries to a QSQ oracle $\mathsf{QStat}_\rho$. In order to simulate $\mathcal{B}$, it's sufficient to simulate each QStat query $(\mathcal{M},\tau)$ by running $\mathcal{A}_{\mathcal{M},\epsilon}$ with parameters $\beta' = \beta/t$ and $m' =  c\cdot\frac{\log(1/\beta')k^2}{\epsilon^2 \tau^2}$ on $m'$ (unused) copies of $\rho$. Then the simulation requires $m'\cdot t $ copies and produces the same output as $\mathcal{B}$ with probability at least $1-\beta$.

The above result generalizes Theorem 6.5 in \cite{arunachalam2020quantum}, as this previous result shows the quantum statistical queries can be simulated by (standard) differentially private measurements. Our result holds under \emph{local} differential privacy, which provides stronger security guarantees, and thus implies the result of \cite{arunachalam2020quantum}. From a practical standpoint, the two results differ as we randomize each outcome $y_i$, while in \cite{arunachalam2020quantum} only the final average is randomized by a single injection of Laplace noise.

\begin{algorithm}

    \textbf{Input}{ $\rho^{\otimes m}$, a $k$-ary POVM $\mathcal{M}$.}\\
    \textbf{Output}{ An estimate of $\mathbb{E}[\mathcal{M}(\rho)]$ up to additive error $\tau$.}
    \begin{enumerate}
        \item Perform the (non-private) measurement $\mathcal{M}$ on each copy of $\rho$ and let $y_1,y_2,\dots, y_m$ be the outcomes.
        \item Sample $\eta_1,\eta_2,...\eta_m$ i.i.d. from the Laplace distribution centered in $0$ and with scale parameter $(k-1)/\epsilon$.
        \item Return $\hat{\mu}:= \frac{1}{m}\sum_{i=1}^m (y_i+\eta_i)$.
    \end{enumerate}
    \label{alg1}
    \caption{A quantum $\epsilon$-LDP algorithm $\mathcal{A}_{\mathcal{M},\epsilon}$ that simulates $\mathsf{QStat}_\rho$}
\end{algorithm}

Combined with the upper bounds provided in \cite{arunachalam2020quantum} and \cite{arunachalam2023role}, \thmref{thm:equiv} readily implies that a wide family of concepts is learnable from quantum examples under local differential privacy, including parities, $k$-juntas, DNF functions and of $n$-qubit trivial states, i.e. the states obtained by applying an arbitrary constant depth circuit to the initial state $\ket{0^n}$.

\subsection{Simulation of locally differentially private measurements with QStat queries}
It remains to show that locally differentially private measurements can be simulated efficiently with quantum statistical queries. We will prove it using a rejection-sampling algorithm, along the lines of Lemma 5.8 in \cite{equiv}.

\begin{theorem}
\label{thm:equiv2}
Let $\mathcal{M}$ be an $\epsilon$-LDP measurement. Then $\mathcal{B}_{\epsilon}$  (\algref{alg2}) in expectation makes $O(e^\epsilon)$ queries to $\mathsf{QStat}_\rho$ with accuracy $\tau = \Theta(\beta/e^{2\epsilon})$ and the total variation distance between $\mathcal{B}_{\epsilon}$'s output distribution and $\mathcal{M}(\rho)$ is at most $\beta$.
\end{theorem}
\begin{proof}
We want to sample from a distribution $\widetilde{p}$ that is within a small total variation distance of $p:= \mathcal{M}(\rho)$.  To this end, we will ensure that, for all $w \in [k]$, $\widetilde{p}(w)$ is a multiplicative approximation of ${p}(w)$. In particular, we show that:
\begin{equation}
\label{eq:approx}
    \widetilde{p}(w) \in (1\pm \phi) p(w) \;\;\;\text{where}\;\;\;\phi=\frac{\beta}{3}.
\end{equation}
Let $\mathcal{M}'$, $q(w)$ and $\tau$ as in \algref{alg2}. Observe the following:
\begin{align}
    \mathbb{E}[\mathcal{M}'(\rho)]= \frac{1 -q(w)}{q(w)(e^\epsilon-e^{-\epsilon})}p(w) -\frac{q(w)}{q(w)(e^\epsilon-e^{-\epsilon})} (1-p(w))
    \\ =\frac{(1 -q(w))p(w) -q(w)(1-p(w))}{q(w)(e^\epsilon-e^{-\epsilon})} = \frac{p(w)-q(w)}{q(w)(e^\epsilon-e^{-\epsilon})} .
\end{align}
Thus,
\begin{equation}
    v = \frac{p(w)-q(w)}{q(w)(e^\epsilon-e^{-\epsilon})} \pm \frac{\beta}{3e^{2\epsilon}}.
\end{equation}
\begin{equation}
    \widetilde{p}(w) = p(w) \left(1 \pm \frac{\beta}{3e^{2\epsilon}}\frac{q(w)}{p(w)}(e^\epsilon - e^{-\epsilon})\right)
\end{equation}
By $\epsilon$-local differential privacy,
\begin{equation}
    e^{-\epsilon}\leq \frac{q(w)}{p(w)} \leq e^{\epsilon}.
\end{equation}
Putting all together we obtain
\begin{equation}
    \widetilde{p}(w)  = p(w) \left(1 \pm \frac{\beta}{3} \right).
\end{equation}

Having established \eqref{eq:approx}, we can show that the algorithm works as desired. First, we notice that the probability on Step 4 is well defined, as \eqref{eq:approx} and $\epsilon$-local differential privacy guarantee that $\frac{\widetilde{p}(w)}{q(w)\left(1+\frac{\beta}{3}\right)}$ is at most 1.
Remark that $\widetilde{p}(w)$ is not a fixed function of $w$: it depends on the QSQ oracle and may vary, for the same $w$, from iteration to iteration. Yet, 
$\widetilde{p}$ is fixed for any given iteration of the algorithm. In the given iteration, any particular element $w$ gets output with probability $q(w) \cdot \frac{\widetilde{p}(w)}{q(w)(1+\phi)e^\epsilon}=\frac{\widetilde{p}(w)}{(1+\phi)e^\epsilon}$.
The probability that the given iteration terminates (i.e., outputs some $w$) is then $p_{terminate} = \sum_w \frac{\widetilde{p}(w)}{(1+\phi)e^\epsilon}$. By \eqref{eq:approx}, this probability is in $\frac{1\pm\phi}{(1+\phi)e^\epsilon}$.
Thus, conditioned on the iteration terminating, element $w$ is output with probability 
\begin{equation}
    \frac{\widetilde{p}(w)}{(1+\phi)e^\epsilon p_{terminate}} \in \frac{1\pm \phi}{1\pm \phi}p(w).
\end{equation}
Since $\phi \leq 1/3$, we can simplify this to get
\begin{equation}
    \Pr[w \text{ output in a given iteration } | \text{ iteration produces output}]\in (1 \pm 3\phi)p(w).
\end{equation}
This implies that no matter which iteration produces output, the total variation distance between the distribution
of $w$ and $p(\cdot)$ will be at most $3\phi = {\beta}$, as desired.
Moreover, since each iteration terminates with probability at least $\frac{1-\phi}{1+\phi} e^{-\epsilon}$, the expected number of
iterations is at most $\frac{1+\phi}{1-\phi}e^\epsilon\leq 2e^\epsilon$. Thus, the total expected QSQ query complexity of the simulation is $O(e^\epsilon)$.
\end{proof}
As for the other direction of the equivalence, also  \thmref{thm:equiv2} can be extended to the case where an algorithm $\mathcal{A}$ accesses $t$ (unused) copies of a state $\rho$ via $\epsilon$-LDP measurements $\mathcal{M}^{(1)}, \mathcal{M}^{(2)},\dots, \mathcal{M}^{(t)}$. In order to simulate $\mathcal{A}$, it's sufficient to simulate each $\epsilon$-LDP measurement $\mathcal{M}^{(i)}$ by running $\mathcal{B}_\epsilon(\mathcal{M}^{(i)})$ with parameters $\beta' = \beta/t$. Then the output distribution of the simulation and the output distribution of $\mathcal{A}$ are within a total variation distance at most $\beta$.
In the classical counterpart of this result \cite{equiv}, the authors provide separate proofs for the adaptive and non-adaptive cases, as they assume that the algorithm $\mathcal{A}$ might reuse some portions of the input dataset. However in our proof we don't need to treat the two cases separately, as we assumed that each measurement is performed on a new copy of the input state $\rho$.
\begin{algorithm}

    \textbf{Input}{ Oracle access to $\mathsf{QStat}_\rho$, $\epsilon\geq 0$, $\beta\geq 0$, an $\epsilon$-LDP measurement $\mathcal{M}=(\mathcal{M}_1,\mathcal{M}_2,\dots,\mathcal{M}_k)$} with outcomes in $[k]$.
    
    \textbf{Output}{ A number $w \sim \widetilde{p}$, such that $\widetilde{p}(w)\in (1\pm \beta/3) p(w)$.}

    \begin{enumerate}
        \item Apply $\mathcal{M}$ to a fixed input, for instance the all-zeros state $\ket{\boldsymbol{0}}:=\ket{00...0}$.  Let $w\sim E(\ket{\bold{0}}\bra{\bold{0}})$ be the outcome.
        \item Define $q(w):=\Tr\{\mathcal{M}_w \ket{\bold{0}}\bra{\bold{0}}\}$ and $\mathcal{M}'=(\mathcal{M}'_0,\mathcal{M}'_1)$, where $\mathcal{M}'_0 :=\mathcal{M}_w$ and $\mathcal{M}'_1:=\mathbb{1}-\mathcal{M}_w$. $\mathcal{M}'_0$ corresponds to the outcome $\frac{1 -q(w)}{q(w)(e^\epsilon-e^{-\epsilon})}$ and $\mathcal{M}'_1$ to the outcome $-\frac{q(w)}{q(w)(e^\epsilon-e^{-\epsilon})}$. 
        Let $\tau = \frac{\beta}{3e^{2\epsilon}}$.
        \item Query the oracle $\mathsf{QStat}_\rho(\mathcal{M}',\tau)$ to compute $v \in \mathbb{E}[\mathcal{M}'(\rho)] \pm \tau$. 
        Define the probability:
        \[
            \widetilde{p}(w) = {vq(w)(e^\epsilon - e^{-\epsilon}) +q(w)}.
        \]
        \item Output $v$ with probability
        \[
             \frac{\widetilde{p}(w) }{q(w)\left(1+\frac{\beta}{3}\right)e^\epsilon}.
        \]
        \item With the remaining probability, repeat from Step 1.
    \end{enumerate}
    \label{alg2}
    \caption{A QSQ algorithm $\mathcal{B}_{\mathcal{M},\epsilon}(\beta,\mathsf{QStat}_\rho)$ that simulates an $\epsilon$-LDP measurement $\mathcal{M}$}
\end{algorithm}

\thmref{thm:equiv2} enables the transfer of lower bounds from the QSQ model to quantum local differential privacy. In particular, (\cite{arunachalam2023role}, Theorem 17) shows that learning the following class in the QSQ model requires exponentially many samples,
\[
\mathcal{C} = \left\{ \ket{\psi_A} = \frac{1}{\sqrt{2^n}}\sum_{x\in\{0,1\}^n} \ket{x,(x^\mathsf{T} A x) \mod 2 : A \in \mathbb{F}^{n\times n}_2}\right\}.
\]
On the other hand, this class is efficiently learnable using separable measurements \cite{arunachalam2022optimal} and entangled measurements with classification noise \cite{arunachalam2023role}. Specifically, this immediately implies an exponential separation between learning under quantum local differential privacy and learning with separable measurements, resolving an open question in (\cite{aaronson2019gentle}, Question 4).

\section{Testing and learning quantum states under local privacy}
We will now explore the effect of local differential privacy in the settings of quantum hypothesis testing and quantum multi-party learning. Intuitively, as local differential privacy is ensured by the injection of noise, this will increase the sample complexity in a testing or learning task. We confirm this intuition by providing a converse bound on the achievable rate for quantum hypothesis testing under local differential privacy. On the other hand, we also demonstrate that quantum local differential privacy is compatible with exponential quantum advantage. As a proof of principle, we prove that parity functions can be learned from quantum examples under local differential privacy.

\subsection{Private hypothesis testing}
\label{sec:hp}
Here we demonstrate an application of the informatic-theoretic results of \secref{sec:entropic} to the rich field of quantum hypothesis testing.
We study the distinguishability of two quantum states $\rho$ and $\sigma$ using a restricted class of measurements, i.e. locally differentially private measurements performed on a single copy of the input state. In particular, we'll consider the task of \emph{asymmetric hypothesis testing}, where one wants
to minimize the rate of false positives (type-1 error) subject to a constraint on the rate of false negatives (type-2 error). 
We will adopt the framework developed by \citet{brandao2014adversarial}, which extends hypothesis testing to the setting of restricted measurements.
Our result can also be regarded as a quantum version of the ``private Chernoff-Stein lemma'' provided in \cite{hockey_stick}.

Let $\rho$ and $\sigma$ be two quantum states acting on some Hilbert space $\mathcal{H}$. Given either $n$ copies of $\rho$ or $n$ copies of $\sigma$, we want to design a test which distinguishes the two possibilities.  For an acceptance operator $\mathcal{M}^n$ (i.e. a POVM element acting on $n$ copies of the input state), we define the error probabilities as follows
\begin{align*}
    \alpha_n(\mathcal{M}^n):=\Tr((I-\mathcal{M}^n)\rho^{\otimes n}) \;\;\;\;\text{(type-2 error)}, \\ \beta_n(\mathcal{M}^n):=\Tr(\mathcal{M}^n\sigma^{\otimes n}) \;\;\;\;\text{(type-1 error)}.
\end{align*}
Then for $0 < \tau < 1$, define
\begin{equation}
\label{eq:beta}
    \beta^\tau_n:=\inf_{\mathcal{M}^n}\{\beta_n(\mathcal{M}^n):\alpha_n(\mathcal{M}^n)\leq \tau\}
\end{equation}
and the asymptotic optimal error exponent 
\begin{equation}
\label{eq:optimal}
    E(\rho,\sigma):= \lim_{\tau \xrightarrow{} 0}\lim_{n\xrightarrow{} \infty} -\frac{\log \beta^\tau_n}{n}.
\end{equation}
The quantum Stein's lemma \cite{hiai1991proper} says that 
\begin{equation}
\label{eq:stein}
    D(\rho\|\sigma) = E(\rho,\sigma).
\end{equation}

As shown by \citet{Ogawa_2005}, the ``strong converse'' \eqref{eq:stein} also holds. This can be thought of as showing that \eqref{eq:stein} is satisfied also when the limit of $\tau \xrightarrow{} 0$ in \eqref{eq:beta} is replaced by any fixed $\tau \in (0, 1)$.
To deal with the restricted case where only single-copy  $\epsilon$-LDP measurements are allowed, we'll need to define the following quantities, introduced in \cite{brandao2014adversarial}.
Consider the infinite set $\bold{S} =(\mathrm{S}^1, \mathrm{S}^2, \dots, \mathrm{S}^n, \dots)$, where each $\mathrm{S}^n$ is a set of measurements over $\mathcal{H}^{\otimes n}$. 
We define:
\begin{equation}
    D_{\mathrm{S}^n}(\rho\|\sigma):=\sup_{\mathcal{M}\in \mathrm{S}^n} \frac{D(\mathcal{M}(\rho^{\otimes n})\|\mathcal{M}(\sigma^{\otimes n}))}{n}.
\end{equation}
\begin{equation}
    D_{\bold{S}}(\rho\|\sigma) := \lim_{n\xrightarrow[]{} \infty}D_{\mathrm{S}^n}(\rho\|\sigma).
\end{equation}
In analogy with \eqref{eq:beta} and \eqref{eq:optimal}, we have
\[
    \beta^\tau_n (\bold{S}):=\inf_{\mathcal{M}\in \mathrm{S}^n}\{\beta_n(\mathcal{M}):\alpha_n\leq \tau\},
\]
\[    E_\bold{S}(\rho,\sigma):= \lim_{\tau \xrightarrow{} 0}\lim_{n\xrightarrow{} \infty} -\frac{\log \beta^\tau_n(\bold{S})}{n}.
\]

We are now ready to upper bound $E_\bold{S}(\rho,\sigma)$ for the case of locally differentially private measurements.

\begin{theorem}[Private quantum Stein's lemma]
\label{thm:stein}
Let $\rho$ and $\sigma$ be two quantum states acting on some Hilbert space $\mathcal{H}$.
Let $\mathrm{S}_\epsilon$ be a set of $\epsilon$-LDP measurements over $\mathcal{H}$. Moreover, for every $n\geq 1$, define the following convex hull
\[
\mathrm{T}^n = \mathsf{conv}\{\mathcal{T}_1\otimes\dots\otimes \mathcal{T}_n : \mathcal{T}_1,\dots,\mathcal{T}_n \in \mathrm{S}_\epsilon\},
\]
and thus let $\bold{T} = (\mathrm{T}^1,\mathrm{T}^2,\dots,\mathrm{T}^n,\dots)$. The following inequality holds:
\[
    E_\bold{T}(\rho,\sigma) \leq \frac{e^\epsilon}{2}(1-e^{-\epsilon})^2 D_M(\rho\|\sigma),
\]
where $D_M(\cdot\|\cdot)$ denotes the measured relative entropy.
\end{theorem}

\begin{proof}
The theorem follows combining the results of \citet{brandao2014adversarial} with \corref{cor:quadratic}. In particular, (\cite{brandao2014adversarial}, Theorem 16) implies that
\[
E_\bold{T}(\rho,\sigma) = D_\bold{T}(\rho\|\sigma).
\]
Recall that
\begin{equation}
\label{eq1}
    D_\bold{T}(\rho\|\sigma) = \lim_{n\rightarrow \infty}\sup_{\mathcal{M}\in \mathrm{T}^n} \frac{D(\mathcal{M}(\rho^{\otimes n})\|\mathcal{M}(\sigma^{\otimes n}))}{n}.
\end{equation}

Observe that $\mathcal{M} = \sum_{i=1}^m {\lambda_i}  (\mathcal{M}_1^{(i)} \otimes\dots \otimes \mathcal{M}_n^{(i)})$ for some non-negative coefficients such that $\sum_i \lambda_i =1$ and $\mathcal{M}_1^{(i)},\dots,\mathcal{M}_n^{(i)} \in \mathrm{S}_\epsilon$. Recall that the quantum relative entropy enjoys joint convexity and additivity with respect to product states. Thus,
\begin{align}
\label{eq2}
\begin{split}
    D(\mathcal{M}(\rho^{\otimes n})\|\mathcal{M}(\sigma^{\otimes n})) = D\left(\sum_{i=1}^m {\lambda_i}  (\mathcal{M}_1^{(i)} \otimes\dots \otimes \mathcal{M}_n^{(i)})(\rho^{\otimes n})\bigg\|\sum_{i=1}^m {\lambda_i}  (\mathcal{M}_1^{(i)} \otimes\dots \otimes \mathcal{M}_n^{(i)})(\sigma^{\otimes n})\right) \\ \leq  \sum_{i,j} \lambda_i  D\left(\mathcal{M}_j^{(i)}(\rho)\bigg\|\mathcal{M}_j^{(i)}(\sigma)\right)\leq n\cdot \max_{i,j} D\left(\mathcal{M}_j^{(i)}(\rho)\bigg\|\mathcal{M}_j^{(i)}(\sigma)\right)
    \\\leq n\cdot\frac{e^\epsilon}{2}(1-e^{-\epsilon})^2 D_M(\rho\|\sigma),
\end{split}
\end{align}
where the last inequality follows directly from \corref{cor:quadratic}. 
Finally, combining \eqref{eq1} and \eqref{eq2} yields
\[
    D_\bold{T}(\rho\|\sigma)\leq \lim_{n\rightarrow \infty} \frac{n}{n}\cdot\frac{e^\epsilon}{2}(1-e^{-\epsilon})^2 D_M(\rho\|\sigma) = \frac{e^\epsilon}{2}(1-e^{-\epsilon})^2 D_M(\rho\|\sigma),
\]

and hence the theorem follows.

\end{proof}

\subsection{Private multi-party learning from quantum data}
\label{sec:parity}
We will now discuss the applications of quantum local differential privacy to the setting of multi-party computation (MPC). In many real-world scenarios, multiple parties share their data to collectively compute a function. The goal is then to achieve the best possible accuracy under some security constraints. One way to formulate the security requirement is to ask that each party learns nothing more
about the other parties' data than can be learned from the output of the function computed. This approach is adopted by the framework of \emph{secure multi-party computation} (SMPC), both in the classical \cite{Yao86, goldreich1987solve} and in the quantum setting \cite{crepeau2002secure, Dulek_2020}. 
The main shortcoming of SMPC is that the security guarantees are dependent on the auxiliary information disposed by the adversary. For instance, if $k$ parties collectively compute an average, $k-1$ malicious parties can collaborate to infer the data of the remaining party. 

To overcome these limitations, we can adopt the framework of \emph{secure multi-party differential privacy}, defined in \cite{kairouz2015secure}.
In particular, we will consider a model where the input state $ \rho_1\otimes \rho_2\otimes\dots\otimes\rho_k$ is distributed among $k$ quantum parties $\mathcal{P}_1,\mathcal{P}_2,\dots,\mathcal{P}_k$, such that the $i$-th party $\mathcal{P}_i$ holds the state $\rho_i$ and disposes of a quantum computer. The parties are allowed to share classical information. 
In order to protect the private information contained in $\rho_i$, we require that the each $\mathcal{P}_i$ accesses the state $\rho_i$ through an  $\epsilon$-local differentially private measurement $\mathcal{M}_i$ for some suitable $\epsilon > 0$.
Thus, for all $i$, for any possible output $y$, and for all input states $\rho_i, \sigma_i$, we have
\[
\Pr[\mathcal{M}_i(\rho_i) = y ] \leq e^\epsilon \Pr[\mathcal{M}_i(\sigma_i) = y ].
\]

One potential concern with this setting is that the injection of noise can severely limit the usefulness of the computation, hence it is no clear a priori whether a quantum speed-up can be achieved under these constraints.
To address this issue, we show that that parity functions can be efficiently learned from quantum examples in a multi-party setting under local differential privacy. Classically, learning parity under local differential privacy requires exponentially many samples \cite{equiv}. 

For $s\in\{0,1\}^n$, the corresponding parity function $c:\{0,1\}^n\rightarrow \{-1,1\}$ is defined $c (x) = (-1)^{s\cdot x}$. Let $b^1,\dots, b^k$ random binary strings in $\{\pm1\}^n$, such that each $b_x^i$ equals $1$ with probability $9/10$ and $-1$ with probability $1/10$. 
Each party $\mathcal{P}_i$ holds the following quantum state:
\begin{equation}
\label{eq:parity}
   \ket{\psi_i} = \sqrt{\frac{1}{2^n}} \sum_{x\in\{0,1\}^n} \ket{x,c(x)\cdot b^i_x}. 
\end{equation}
We remark that this definition slightly differs from the one considered in (\cite{arunachalam2020quantum}, Lemma 4.2), as their definition doesn't involve the random vector $b_i$.
Instead, in our model each party holds a different input state. {The vector $b^i$ can be either regarded as classification noise or as some sensitive information regarding the $i$-th party. In the latter case, the adoption of local differential privacy is extremely natural, as it significantly limits the information about $b^i$ that can be inferred by a malicious adversary, even disposing of auxiliary information}. 
\begin{prop}
Let $s\in\{0,1\}^n$ and $\ket{\psi_1}, \ket{\psi_2},\dots, \ket{\psi_k}$ as defined above and assume that the parties $\mathcal{P}_i$'s can communicate via a classical channel. Provided that  $k\geq c \cdot n\epsilon^{-2}\log(1/\beta)$ for a sufficiently large constant $c$, there is an efficient quantum algorithm $\mathcal{A}$ that computes the string $s$ with probability at least $1-\beta$. $\mathcal{A}$ consists solely in $\epsilon$-LDP measurements on the states $\ket{\psi_i}$'s, classical communication and classical post-processing.
\end{prop}

\begin{proof}
The proof is similar to the one of (\cite{arunachalam2020quantum}, Lemma 4.2). It is not hard to see that $\mathrm{Inf}_j(c) = 1$  for all $j \in \mathrm{supp}(s)$ and $\mathrm{Inf}_j(c) = 0$ otherwise. 
As shown in \cite{arunachalam2020quantum}, there is a quantum measurement $M_j$ implementable in $\poly(n)$ gates such that
\[
\bra{\psi}M_j \ket{\psi} = \mathrm{Inf}_j(c),
\]
where $\ket{\psi} = \sqrt{\frac{1}{2^n}} \sum_{x\in\{0,1\}^n} \ket{x,c(x)}$.
Moreover, the expected trace distance between $\ket{\psi}$ and $\ket{\psi_i}$ can be bounded as follows:
\[
\mathbb{E}_{b_i}\|\ket{\psi}\bra{\psi} - \ket{\psi_i}\bra{\psi_i}\|_\tr = \mathbb{E}_{b_i} \left[\sqrt{1 - \braket{{\psi}_i}{\psi}} \right] = \sqrt{1- \sqrt{1-1/10}} < \frac{1}{4},
\]
where we took the expectation over the randomness of the string $b_i$. By the property of the trace distance,
\[
\left|\mathbb{E}_{b_i}\bra{\psi_i}M_j \ket{\psi_i}  - \mathrm{Inf}_j(c) \right|< \frac{1}{4}.
\]
Then the algorithm $\mathcal{A}$ estimates $\mathrm{Inf}_j(c)$ by asking $m > 64\cdot \epsilon^{-2}\log(3/\beta)$ parties to perform a Laplace measurement $M_j^{\mathrm{Lap}, \epsilon}$ on their state $\ket{\psi_i}$ and averaging the outcomes $\hat{y}_1,\hat{y}_2,\dots,\hat{y}_m$. We denote their average by $\hat{\mu} = \frac{1}{n} \sum_{i=1}^m \hat{y}_i$. We can write $\hat{y}_i = y_i + \eta$, where   $y_i \sim \mathbb{E}_{b_i}\bra{\psi_i}M_j \ket{\psi_i}$ and $\eta \sim \mathrm{Lap}(1/\epsilon)$.
Proceeding as in the proof of \thmref{thm:equiv}, we can show by concentration of measure that 
\[
\hat{\mu} = \mathbb{E}_{b_i}\bra{\psi_i}M_j \ket{\psi_i} \pm 1/4 ,
\]
with probability at least $1- \beta$. 
Then the outcome $\hat{\mu}$ is in the interval $(1/2,3/2)$ if $j \in \mathrm{supp}(s)$, otherwise is in $(-1/2,1/2)$. Thus we can determine whether $j \in \mathrm{supp}(s)$.
Repeating the procedure for all $j \in [n]$ on $m$ unused states $\ket{\psi_i}$'s, we can determine the string $s$. This requires $k$ to scale as $O( n\epsilon^{-2}\log(1/\beta))$.

\end{proof}

\section*{Acknowledgments}
The authors thank Daniel Stilck-França, Christoph Hirche, Alex B. Grilo and Mina Doosti for helpful discussions at different stages of this project.
AA acknowledges financial support from the QICS (Quantum Information Center Sorbonne). AA and EK acknowledge financial support from the H2020-FETOPEN Grant PHOQUSING (GA no.: 899544).


\bibliography{sample}
\bibliographystyle{unsrtnat}

\appendix
\section{The measured Pinsker's inequality}
We provide an alternative version of the popular Pinsker's inequality \cite{hiai1981sufficiency}, where the quantum relative entropy is replaced by the measured relative entropy. As the proof is almost identical to the one of the (standard) quantum Pinsker's  inequality, this can be regarded as a folklore result.
We include it here since we were unable to find an appropriate reference.

\begin{lemma}[Measured Pinsker's inequality]
\label{lem:pinsker}
For $\rho,\sigma$ quantum states, the following inequality holds:
\[
\|\rho-\sigma\|_\tr^2 \leq \frac{1}{2} D_M(\rho\|\sigma),
\]
where $D_M(\rho\|\sigma)$ is the measured relative entropy.
\end{lemma}
\begin{proof}
Recall the variational interpretation of the trace distance as a probability difference:
\[
\|\rho-\sigma\|_\tr = \max_{0\leq \Lambda \leq \mathbb{1}} \Tr[\Lambda(\rho-\sigma)] = \stat{\mathcal{M}(\rho) - \mathcal{M}(\sigma)},
\]
where $\mathcal{M}=(\Lambda^*, \mathbb{1} - \Lambda^*)$ and $\Lambda^* = \arg\max_{0\leq \Lambda \leq \mathbb{1}} \Tr[\Lambda(\rho-\sigma)] $.
The classical Pinsker's inequality yields:
\[
\stat{\mathcal{M}(\rho) - \mathcal{M}(\sigma)}^2 \leq \frac{1}{2} D(\mathcal{M}(\rho)\|\mathcal{M}(\sigma))\leq \frac{1}{2} D_M(\rho\|\sigma),
\] 
where the second inequality follows from the definition of measured relative entropy. This proves the lemma.
\end{proof}
The standard inequality can be deduced by noting that $D_M(\rho\|\sigma)\leq D(\rho\|\sigma)$.

\end{document}